\documentclass[11pt]{article}
\usepackage{amsmath}
\usepackage{amssymb}
\usepackage{epsfig}
\newcounter{thmc}
\newcounter{defc}

\newtheorem{theorem}[thmc]{Theorem}

\newtheorem{definition}[defc]{Definition}
\newtheorem{lemma}{Lemma}
\newcommand{\OL} {\overline}

\begin{document}
\title{Algorithms for Unipolar and Generalized Split Graphs}
\author{Elaine M. Eschen
\thanks{
elaine.eschen@mail.wvu.edu, Lane Dept. of Computer Science and
Electrical Engineering, West Virginia University, Morgantown, WV
26506. } \and Xiaoqiang Wang
\thanks{
xiaoqiang.wang.wvu@gmail.com, Lane Dept. of Computer Science and
Electrical Engineering, West Virginia University, Morgantown, WV
26506. }} \maketitle

\begin{abstract}
A graph $G=(V,E)$ is a {\it unipolar graph} if there exists a
partition $V=V_1 \cup V_2$ such that, $V_1$ is a clique and $V_2$
induces the disjoint union of cliques. The complement-closed class
of {\it generalized split graphs} contains those graphs $G$ such that
either $G$ {\it or} the complement of $G$ is unipolar. Generalized
split graphs are a large subclass of perfect graphs. In fact, it has
been shown that almost all $C_5$-free (and hence, almost all perfect
graphs) are generalized split graphs. In this paper we present a
recognition algorithm for unipolar graphs that utilizes a minimal
triangulation of the given graph, and produces a partition when one
exists. Our algorithm has running time O($nm+nm_F$), where $m_F$ is
the number of edges added in a minimal triangulation of the given
graph. Generalized split graphs can be recognized via this algorithm
in O($n^3$) time. We give algorithms on unipolar graphs for finding
a maximum independent set and a minimum clique cover in O($n+m$)
time and for finding a maximum clique and a minimum proper coloring
in O($n^{2.5}/\log n$) time, when a unipolar partition is given.
These algorithms yield algorithms for the four optimization problems
on generalized split graphs that have the same worst-case time
bounds. We also report that the perfect code problem is NP-Complete
for chordal unipolar graphs.
\end{abstract}


\section{Introduction}
The graphs in this paper are finite, simple, and undirected.  For
graph-theoretic terms not defined here and well-known graph theory
concepts see \cite{BM}.

The class of polar graphs, introduced by Tyshkevich and Chernyak
\cite{TCdecomp} in 1985, has received a lot of attention recently. A
\emph{complete multipartite graph} is the complement of a disjoint
union of complete graphs. A graph $G$ is \emph{polar} if its vertex
set can be partitioned into two sets $A$ and $B$ such that the
subgraph induced by $A$ in $G$ is a complete multipartite graph and
the subgraph induced by $B$ in $G$ is the complement of a complete
multipartite graph (i.e., the disjoint union of complete graphs).
When $A$ is restricted to be an independent set, $G$ is said to be
\emph{monopolar}; when $A$ is restricted to be a clique, $G$ is said
to be \emph{unipolar}. Unipolar graphs have also been called
\emph{clique-split} graphs in the literature \cite{S03}.
A graph is a \emph{split graph} if its vertex set can be partitioned
into an independent set and a clique. The complement of a split
graph is also a split graph. Polar graphs generalize bipartite and
split graphs. If one replaces one of the bipartitions of a bipartite
graph with a disjoint union of cliques, then a monopolar graph is
obtained. If one replaces the independent set of a split graph with
a disjoint union of cliques, then a unipolar graph is obtained.

If the complement of a graph $G$ belongs to a given class
$\mathcal{C}$, then we say that $G$ is {\it co-$\mathcal{C}$}. For
example, a graph whose complement is unipolar is said to be
co-unipolar. The class of \emph{generalized split graphs} is equal
to the union of the class of unipolar graphs and the class of
co-unipolar graphs. The graph classes unipolar, co-unipolar, and
generalized split graphs are hereditary. The class of polar graphs
is closed under complementation, while the complement of a monopolar
(resp., unipoloar) graph is polar but not necessarily monopolar
(resp., unipolar). Note that the class of generalized split graphs
is closed under complementation.

The recognition problem for a graph class $\mathcal{C}$ is as follows: Given a
graph $G$, is $G$ a member of class $\mathcal{C}$? The recognition
problem for both polar graphs \cite{CC} and monopolar graphs
\cite{AF} has been shown to be NP-Complete.  Recent research is
focused on the polar and monopolar recognition problems on special
classes of graphs. Churchley and Huang \cite{CHclaw} prove that
testing for polarity remains NP-Complete for claw-free graphs and
testing for monopolarity remains NP-Complete for triangle-free
graphs, while showing that monopolarity can be decided efficiently
for claw-free graphs. Le and Nevries \cite{LN,LNplanar} extend
NP-Completeness results for both recognition problems to several
special classes of graphs, and establish that hole-free, $P_5$-free,
($2K_2,C_5$)-free, and chair-free graphs have polynomial-time
monopolarity tests, while polarity is NP-Complete for these classes.
Polynomial-time algorithms for recognizing polar and monopolar
graphs are known for a host of other special classes of graphs,
including cographs \cite{EMW}, chordal graphs \cite{EHSW},
permutation graphs \cite{EHM}, line graphs \cite{CHline,EH,HX}, and
graph classes with bounded tree-width \cite{ALS,BC} or bounded
clique-width \cite{BMR}.

Tyshkevich and Chernyak \cite{TCpolarity} showed that unipolar
graphs can be recognized in O($n^3$) time.  Churchley and Huang
\cite{CHcolor} give an alternate O($n^2m$)-time algorithm using a
reduction to a polynomial-time solvable 2-edge-colored homomorphism
problem. In this paper we present a recognition algorithm for
unipolar graphs that utilizes a minimal triangulation of the given
graph. Our algorithm has running time O($nm+nm_F$), where $m_F$ is
the number of edges added in a minimal triangulation of the given
graph. Thus, our algorithm may be more efficient, and in all cases
is not less efficient, than the O($n^3$) algorithm.  Also, it is
more efficient than the O($n^2m$) algorithm when $m_F$ is
sufficiently small. When our unipolar recognition algorithm is given
a graph in the class, the algorithm produces a unipolar partition in
O($nm+nm_F$) time. Generalized split graphs can be recognized in
O($n^3$) time via the unipolar recognition algorithm of Tyshkevich
and Chernyak \cite{TCpolarity} or via the unipolar recognition
algorithm presented in this paper (note we may have to work on the
complement of the given graph).

We denote the chordless cycle on $k$ vertices by $C_k$. We shall
call a chordless cycle on five or more vertices a {\it hole} and the
complement of a chordless cycle on five or more vertices an {\it
antihole}. Holes and antiholes are designated even or odd depending
on whether they have an even or odd number of vertices.

Generalized split graphs can contain even holes, $C_4$, and even
antiholes (since an even antihole can be partitioned into two
cliques). It is not difficult to see that holes and odd antiholes
are not unipolar graphs (see Lemmas \ref{obs-no-holes} and
\ref{obs-no-odd-aholes}). Thus, unipolar, co-unipolar, and
generalized split graphs are perfect via the Strong Perfect Graph
Theorem \cite{CRST}, which states that a graph is \emph{perfect} if and
only if it does not contain an odd hole or an odd antihole as an
induced subgraph.

A graph is {\it chordal} if it does not contain an induced cycle on
four or more vertices. F\"{o}ldes and Hammer \cite{FH} proved that
the class of split graphs is equivalent to the class of graphs that
are both chordal and co-chordal. The class of split graphs is
properly contained in the intersection of unipolar and co-unipolar;
a $P_5$, for instance, is not a split graph, but is both unipolar
and co-unipolar. However, the class of generalized split graphs is
incomparable to both the classes of chordal and co-chordal graphs.
The graph $G_c$ consisting of a triangle and a $P_5$ joined by the
single edge between a vertex of the triangle and an endpoint of the
$P_5$ is chordal but not generalized split. The complement of $G_c$
is co-chordal but not generalized split. A graph is \emph{weakly
triangulated} if contains neither a hole nor an antihole. Weakly
triangulated graphs are a well-known class of perfect graphs that
generalize chordal graphs and co-chordal graphs. Along with the fact
that generalized split graphs can contain even holes and even
antiholes, $G_c$ also establishes that the class of generalized
split graphs is incomparable to the class of weakly triangulated
graphs. It is easy to find graphs in the intersection of generalized
split with chordal, co-chordal and weakly chordal graphs. A
superclass of the class of perfect graphs is the class of $C_5$-free
graphs. A graph is $C_5$-free if it does not contain an induced
cycle of length 5.

The class of generalized split graphs was introduced by Pr\"{o}mel
and Steger \cite{PS} in their probabilistic study of perfect graphs.
Let $GS(n)$ denote the set of all labeled generalized split graphs
on $n$ vertices, $P(n)$ denote the set of all labeled perfect graphs
on $n$ vertices, and $F(n)$ denote the set of all labeled $C_5$-free
graphs on $n$ vertices. Pr\"{o}mel and Steger prove the following
theorem, which provides a structural characterization of {\it almost
all} $C_5$-free graphs.

\begin{theorem} {\normalfont (Pr\"{o}mel and Steger \cite{PS})}
Almost all $C_5$-free graphs are generalized split graphs in the sense that
$|GS(n)|/|F(n)| \to 1$, as $n\to\infty$.
\end{theorem}

Since $GS(n) \subset P(n) \subset F(n)$ this theorem implies that
almost all perfect graphs are generalized split graphs. A
consequence of this theorem is that properties established for
generalized split graphs are immediately properties of almost all
$C_5$-free (and almost all perfect) graphs. Bacs\'{o} et al.
\cite{BGGPS} employ this technique to show that the
clique-hypergraphs of almost all perfect graphs are 3-colorable.

Szwarcfiter and Maffray \cite{S03} posed the problem of solving
optimization problems on unipolar graphs and generalized split
graphs. In this paper we consider the unweighted (cardinality)
versions of four classical optimization problems. We give
O($n+m$)-time algorithms to find a maximum independent set and a
minimum clique cover in a unipolar graph when a unipolar partition
is given. We also give O($n^{2.5}/\log n$)-time algorithms to find a
maximum clique and a minimum proper coloring in a unipolar graph
when a unipolar partition is given. These algorithms yield
algorithms for the four optimization problems on generalized split
graphs that have the same worst-case time bounds. If a unipolar
partition is not given as input, finding a unipolar partition of the
input graph can dominate the running time.  These four optimization
problems are NP-Complete for arbitrary graphs (and even for many
special graph classes). However, both the unweighted and weighted
versions are solvable in polynomial time on perfect graphs due to a
result of Gr\"{o}tschel, Lov\'{a}sz, and Schrijver \cite{GLS}.  This
result uses the ellipsoid method for convex programming and the
algorithm is difficult. Furthermore, no other polynomial-time method
for solving these problems on perfect graphs is known.  Hence, there
is interest in simpler efficient combinatorial algorithms to solve
these problems on subclasses of perfect graphs.

\section{Preliminaries}

Let $G = (V,E)$ be a finite, simple, undirected graph, with $|V|=n$
and $|E|= m$. We use $\OL{G} = (V,\OL{E})$ to represent the
complement of $G$, with $|\OL{E}| = \OL{m}$. We use $G_W$ to denote
the subgraph induced in $G$ by $W \subseteq V$. A subset $H
\subseteq V$ is a {\it clique} in $G$ if $G_H$ is a complete graph.
A clique $H$ is {\it maximal} if there is no clique of $G$ that
properly contains $H$ as a subset.

A {\it triangulation} of a given graph $G=(V,E)$ is an embedding of
$G$ in a chordal (triangulated) graph $G^\prime=(V, E \cup F)$ by
adding the set of edges $F$ to $G$. If $F$ is inclusion minimal,
then the triangulation is said to be {\it minimal} and the resulting
chordal graph is called a {\it minimal triangulation of G}. We use
$m_F$ to denote $|F|$ and $m'$ to denote $|E \cup F|$. Similarly,
the number of edges added in a minimal triangulation of $\OL{G}$ is
denoted by $\OL{m}_F$.

We will use the following definition.

\begin{definition}
A graph $G=(V,E)$ is \emph{unipolar} if $V$ can be partitioned into
one or more cliques $H, H_1,...,H_k$ such that there are no edges
between the vertices of $H_i$ and $H_j$ when $i\not=j$ (adjacencies
between vertices of $H$ and vertices of cliques $H_i$ are
arbitrary). The collection of cliques $H, H_1, \ldots, H_k$ is said
to be a \emph{unipolar partition} of $G$ with \emph{center} $H$ and
\emph{peripheral} cliques $H_1, \ldots, H_k$.
\end{definition}

For unipolar graphs, we have the following lemmas.

\begin{lemma}\label{obs-no-holes}
 If $G$ is unipolar, then $G$ contains no hole as an induced subgraph.
\end{lemma}
\textbf{Proof.}
Let $H, H_1, \dots, H_k$ be an arbitrary unipolar partition of $G$.
First, a hole cannot be contained in a single clique. Now suppose
$C$ is a cycle that intersects two distinct peripheral cliques $H_i$
and $H_j$. Since $H$ is a cut set, $C$ has at least two vertices in
$H$, and there is a chord. Now suppose that a cycle $C$ is contained in
$G_{H \cup H_i}$ for some $i$ and that the length of $C$ is at least
5. Then either $H$ or $H_i$ contains at least three vertices of
$C$, and a chord is created. \hfill $\Box$

\begin{lemma}\label{obs-no-odd-aholes}
 If $G$ is unipolar, then $G$ contains no odd antihole as an induced subgraph.
\end{lemma}
\textbf{Proof.} Let $H, H_1, \ldots, H_k$ be an arbitrary unipolar
partition of $G$, and consider an antihole in $G$. First, an
antihole cannot be contained in a single clique. Then, since an
antihole is connected, some vertex $x$ of the antihole must be in
the center $H$. In the antihole $x$ has two adjacent non-neighbors
$y$ and $z$, and these must be in the same peripheral clique $H_i$.
Now all other vertices of the antihole are adjacent to $y$ or $z$
(or both), and therefore cannot be in any peripheral clique $H_j, j
\neq i$. Thus, the antihole is contained in $G_{H \cup H_i}$. The
complement of $G_{H \cup H_i}$ is bipartite, which implies that the
antihole must be even. \hfill $\Box$

\begin{lemma}\label{obs-C4}
If $G$ is unipolar and contains an induced $C_4$, then in any
unipolar partition of $G$, $H, H_1, ..., H_k$, the vertices of the
$C_4$ can be labeled $a,b,c,d$ so that edge $a-b$ is in $H_j$, for
some $j$, and edge $c-d$ is in $H$.
\end{lemma}
\textbf{Proof.} Following from the proof of Lemma
\ref{obs-no-holes}, an induced cycle of length 4 must be
contained in $G_{H \cup H_i}$ for some $i$. If either $H$ or $H_i$
has at least three vertices of this cycle, then there must be a
chord, since both $H$ and $H_i$ are cliques. \hfill $\Box$

\begin{lemma}\label{obs-triang}
If $G=(V,E)$ is unipolar and $G^\prime=(V, E \cup F)$ is a minimal
triangulation of $G$, then, in any unipolar partition $H, H_1, ...,
H_k$ of $G$, the endpoints of edges in $F$ will be such that one is
in $H$ and the other is in $H_j$, for some $j$.
\end{lemma}
\textbf{Proof.}
The proof follows from Lemmas \ref{obs-no-holes} and \ref{obs-C4}, and the fact that
the triangulation is minimal.
\hfill $\Box$

Lemma \ref{obs-triang} also follows from Lemma \ref{obs-C4} and the
following characterization due to Rose, Tarjan, and Lueker
\cite{RTL}.
\begin{theorem} \normalfont{\cite{RTL}}
A triangulation $G^\prime$ is minimal if and only if every fill edge
is the unique chord of a $C_4$ in $G^\prime$.
\end{theorem}

\section{Recognition Algorithm}

Let $G=(V,E)$ be an arbitrary undirected graph. Let $G^\prime=(V,E
\cup F)$ be an arbitrary minimal triangulation of $G$.

\begin{definition}
\normalfont Let $G'$ have a unipolar partition with center clique
$H'$ and peripheral cliques $H'_1, H'_2, \ldots, H'_k$. A
\emph{transferable set} $S$ (if it exists) is a subset of $H'$ such
that, for some $i$, $S \cup H'_i$  is a clique in $G$ and the
vertices of $S$ are independent of the vertices in all sets $H'_j$,
$j\not=i$.
\end{definition}

\begin{definition}\label{feasible}
\normalfont A unipolar partition $H', H'_1, \ldots, H'_k$ of $G'$ is
said to be \emph{feasible} under
the following conditions: \\
i) each edge of $F$ either has one endpoint in $H'$ and the other in $H'_j$, for some $j$,
or both endpoints in $H'$, and\\
ii) if there are edges of $F$ with both endpoints in $H'$, then there exists a transferable
set $S \subseteq H'$ such that all such edges have one endpoint in $H'-S$ and the other in $S$.
\end{definition}

Our recognition algorithm for unipolar graphs is based on the
following theorem.

\begin{theorem} \label{main}
$G$ is unipolar if and only if $G'$ has a feasible unipolar
partition with center $H'$ that is a maximal clique of $G'$.
\end{theorem}
\textbf{Proof.} If $G$ is a complete graph, then the theorem is
trivially true. So suppose that $G$ is not a clique.

\noindent ($\Rightarrow$) Suppose that $G$ has a unipolar partition $H,
H_1, \ldots, H_k$. If $H$ is not a maximal clique, then one or more
vertices from exactly one $H_i$ can be added to $H$ so that the
resulting set is a maximal clique. So assume $H$ is a maximal clique
of $G$. Lemma~\ref{obs-triang} tells us exactly how the edges of $F$
are placed with respect to the unipolar partition of $G$. If $H$ is
also a maximal clique of $G'$, we are done. Otherwise, a subset $S$
of vertices from exactly one $H_j$ can be added to $H$ to form a
maximal clique $H'$ of $G'$. The resulting unipolar partition of
$G'$ with center $H'$ is feasible with transferable set $S$.

\noindent ($\Leftarrow$) Suppose that $G'$ has a feasible unipolar
partition with center $H'$ that is a maximal clique of $G'$. If each
edge of $F$ has one endpoint in $H'$ and the other in $H'_j$, for
some $j$, then the unipolar partition of $G'$ is also a unipolar
partition of $G$. Otherwise, $G'$ has a transferable set $S
\subseteq H'$ satisfying ii in Definition~\ref{feasible}. A unipolar
partition of $G$ can be obtained by transferring $S$ to the
appropriate clique $H'_i$. \hfill $\Box$

If a disconnected graph $G$ is unipolar, then the center of the
unipolar partition must be contained in a single component of $G$
and all other components must be complete graphs. If the complement
of a graph $G$ is bipartite, then $G$ can be covered by two cliques,
and hence is unipolar.
For a vertex $x$ and a set of vertices $H$,
we use {\it $x$ sees $H$} to mean that $x$ is adjacent
to each vertex of $H$.

\noindent
{\bf Algorithm Unipolar\_Test}\\
{\bf Input:} An arbitrary undirected graph $G$. $G$ may be disconnected.\\
{\bf Output:} YES and a unipolar partition of $G$ or NO (i.e., $G$
has no unipolar partition).

\begin{enumerate}
\item Find the components of $G$.
If all components are complete, then a unipolar partition has been
found; stop with YES. If more than one component is not complete,
stop with NO. Otherwise, set $G$ to be the one component that is not
complete and find a unipolar partition of this subgraph.
\item Test whether $\OL{G}$ is bipartite; if so, stop with YES.
\item Find a minimal triangulation $G'$ of $G$.
\item Generate the maximal cliques of $G'$.
\item For each maximal clique $H'$ of $G'$, test whether $H'$ is the center clique of a
feasible unipolar partition of $G'$. If any such center is found,
stop with YES. Otherwise, $G$ is not unipolar
by Theorem \ref{main}.\\

{\bf Details of Step~5:}
\begin{enumerate}
\item Check that the components of $G-H'$ are complete graphs.
If no, $H'$ cannot be the center of a feasible unipolar partition of
$G'$; repeat Step~5 with the next maximal clique of $G'$. If yes,
then the edges of $F$ satisfy condition i in the definition of a
feasible unipolar partition (Definition \ref{feasible}). The
components of $G-H'$ are now the peripheral cliques $H'_i$ of a
candidate feasible unipolar partition of $G'$.
\item If there are no edges of $F$ with both endpoints in $H'$, then (with (a))
$H', G-H'$ is a unipolar partition of $G$; stop with YES.
\item If there are edges of $F$ with both endpoints in $H'$,
attempt to construct the desired transferable set $S$.
If $S$ is found, stop with YES.  If $S$ not found, repeat Step~5 with the next maximal clique
of $G'$. \\

{\bf Details of Step~5(c):}\\

Let $F^*$ be the edges of $F$ with both endpoints in $H'$, and let $V^*$ be the
set of endpoints of edges in $F^*$.
If $S$ exists, then for each edge in $F^*$, exactly one endpoint is in $S$.
\begin{enumerate}
\item For each vertex $v$ of $V^*$ determine whether $v$ sees in $G$
exactly one peripheral clique $H'_i$ and has no neighbors in any
other peripheral clique. Vertices that fail this test cannot be in
$S$; remove these from $V^*$ and place them in $V^-_1$.
\item If any edge of $F^*$ has both endpoints in $V^-_1$, then there is no transferable set;
go to Step~5. Otherwise, go to Step~iii.
\item Choose an arbitrary edge $x-y$ in $F^*$, where $x$ is in $V^*$.
We first try to build a transferable set $S$ containing $x$. Vertex
$x$ sees exactly one peripheral clique $H'_i$ and $S$ must be
transferable to this clique. Now vertices that do not see $H'_i$
cannot be in $S$; remove these from $V^*$ and add them to $V^-_2$.
If a transferable set is found, stop with YES. If a transferable set
is not found and $y$ is in $V^-_1$, go to Step~5. Otherwise, attempt
to build a transferable set $S$ containing $y$. Restore $V^*$ by
setting $V^* = V^* \cup V^-_2$, and then set $V^-_2 = \emptyset$.
If successful, stop with YES; else, go to Step~5.\\

{\bf Details of finding a transferable set $S$ containing a vertex $v$:}\\

This can be reduced to an instance of 2-Satisfiability. Each vertex
of $V^* \cup V^-_1 \cup V^-_2$ corresponds to a variable. A vertex
is in $S$ if and only if its corresponding variable is assigned the
truth value TRUE.
\begin{itemize}
\item For the vertex $v$, construct the clause ($v+v$).  This asserts that $v$ must be in $S$.
\item For each vertex $u$ in $V^-_1 \cup V^-_2$, construct the clause ($\OL{u}+\OL{u}$).
These clauses assert that these vertices cannot be in $S$.
\item For each edge $a-b$ in $F^*$, construct the clauses ($a+b$) and ($\OL{a}+\OL{b}$).
These clauses assert that for each edge in $F^*$ exactly one of its endpoints
is in $S$.
\end{itemize}
Clearly, this Boolean formula is satisfiable if and only if there exists a transferable
set $S$ containing vertex $v$.
\end{enumerate}
\end{enumerate}
\end{enumerate}

Step~1 requires O($n+m$) time. From this point on, we can assume
that graph we are working with is connected; that is, the number of
edges is at least the number of vertices minus 1.
Step~2 can be done in O($n+\OL{m}$) time. A minimal triangulation of
$G$ can be computed in O($nm$) time \cite{AB,ABHSV,RTL}. The number
of maximal cliques of a chordal graph is no more than the number of
vertices \cite{FG}. Furthermore, the maximal cliques can be listed
in linear time (see \cite{Gol,RTL}). Thus, Step~4 requires
O($n+m'$)= O($m+m_F$) time. The 2-Satisfiability algorithm of
reference \cite{2sat} both tests the satisfiability of a 2-CNF
formula and finds a satisfying assignment, when one exists, in time
linear in the size of the formula (where the size is the number of
variables plus the number of clauses). The size of the formula
constructed in Step~5(c)iii is O($n+m_F$). The other parts of Step~5
can be accomplished in O($m+m_F$) time. Hence, Step~5 requires
O($m+m_F$) time. Thus, the overall time required by algorithm
Unipolar\_Test is O($nm+nm_F$).

To determine whether a given graph $G$ is a generalized split graph,
we first test whether $G$ is unipolar; if not, the test is repeated
on $\overline{G}$. In the worst case we run algorithm Unipolar\_Test
on $G$ and $\OL{G}$; thus, we can recognize generalized split graphs
in O($nm + nm_F + n\OL{m} + n\OL{m}_F$) = O($n^3$) time.

\section{Optimization Algorithms}

\subsection{Maximum Independent Set and Maximum Clique}
\label{maxopt}

Let $G=(V,E)$ be a unipolar graph that has a unipolar partition $H,
H_1, \ldots, H_k$. We first discuss finding a maximum independent
set and maximum clique in $G$. Then we present algorithms for
generalized split graphs.

A maximum independent set in $G$ is easily found. An independent set
in $G$ can contain at most one vertex from each clique of the split.
Thus, the size of a maximum independent set in $G$ is either $k+1$
or $k$, depending on whether or not there is a vertex $x \in H$ that
has a non-neighbor $y_i$ in each clique $H_i$. This can be
determined in O($n+m$) time by counting the number of neighbors each
vertex has in each $H_i$ in which it has a neighbor. If such a
vertex $x$ exists, then $G$ has the maximum independent set $\{x,
y_1, \ldots, y_k\}$. The vertices $y_i$ can be found by a single
scan of the vertex set after marking the vertices that are neighbors
of $x$. If no such vertex exists, then $G$ has a maximum independent
set consisting of exactly one arbitrarily chosen vertex from each
clique $H_i$.

A maximum clique in $G$ must be a subset of $H \cup H_i$, for some $i$.
To find a maximum clique in $G$ we determine the size of a maximum clique in $G_{H \cup H_i}$,
for each $i$, and construct a maximum clique in the subgraph that has largest maximum clique.
However, the complement of
$G_{H \cup H_i}$, for each $i$, is a bipartite graph with bipartition $H, H_i$;
thus, we can instead investigate maximum
independent sets in these bipartite graphs.

The K\"{o}nig-Egerv\'{a}ry Theorem \cite{EE,DK} establishes a close
relationship between maximum matchings and minimum vertex covers in
a bipartite graph; namely, in a bipartite graph the size of a
maximum matching equals the size of a minimum vertex cover. Hence,
if $M$ is a maximum matching in $G$, then $|V| - |M|$ is the size of
a maximum independent set in $G$. Furthermore, the proof of this
theorem shows that given a maximum matching in a bipartite graph, a
simple linear time search strategy can be used to find a maximum
independent set (see \cite{BM}). Therefore, a maximum clique of
$G_{H \cup H_i}$ can be found by running a bipartite matching
algorithm on $\OL{G}_{H \cup H_i}$ to obtain a maximum independent
set in $\OL{G}_{H \cup H_i}$.

Thus, to find a maximum clique in $G$,
we solve $k$ bipartite matching problems in $\OL{G}$.
Note that 
we only need to find the size of the maximum independent set
in each of the subgraphs $\OL{G}_{H \cup H_i}$
(i.e., we are looking for the subproblem with the smallest maximum matching)
and then do one O($n+\OL{m}$)-time search
to construct a maximum independent set in $\OL{G}$.

The best-known algorithms for maximum matching in a bipartite graph
are the O($m\sqrt{n}$)-time algorithm of Hopcroft and Karp
\cite{HK}, an O($n^{1.5}\sqrt{m/\log n}$)-time algorithm due to Alt
et al. \cite{ABMP}, and an O($n^{2.5}/\log n$)-time algorithm due to
Feder and Motwani \cite{FM}. The latter two algorithms are more
efficient than the Hopcroft and Karp algorithm when $m$ is
$\Omega(n^2/\log n)$.

To analyze the worst-case time for solving the $k$ bipartite
matching problems in $\OL{G}$, suppose that there are $x_0n$ vertices in
the center set and $x_in$ vertices in the $i$th peripheral set, for
$i=1,2,\ldots, k$, where $\Sigma_{i=0 \,\mathrm{to}\, k}\,x_i = 1$.
In the $i$th problem we have $(x_0+x_i)n$ vertices.
For the Feder-Motwani algorithm the total running time for all $k$
problems is O($\Sigma_{i=1 \,\mathrm{to}\,
k}\,(n^{2.5}(x_0+x_i)^{2.5})/\log({n(x_0+x_i)})$). For fixed $k$,
using elementary techniques (such as the Lagrange multiplier
method), it can be shown that the maximum value of this sum is when
one of the $x_i$ values is as large as possible and the others are
equal to zero; say, $x_1 = 1-x_0$ and $x_i=0$ for $i > 1$. So the
worst case is when $k=1$. Thus, the worst-case time is the time to
solve a single bipartite matching in $\OL{G}$, which is
O($n^{2.5}/\log n$).

To find a maximum independent set in a generalized split graph $G$,
we use a unipolar partition of $G$ or $\OL{G}$. If a unipolar
partition is not given as input, we use algorithm Unipolar\_Test to
obtain a unipolar partition of $G$, if possible; otherwise, we find
a unipolar partition of $\OL{G}$. In this case the time to find the
unipolar partition, which is O($nm + nm_F$) or O($n\OL{m} +
n\OL{m}_F$), can dominate the time to find a maximum independent
set.
\begin{description}
\item[Case 1.]
If $G$ is unipolar, we find a maximum independent set in $G$. This
can be done in O($n + m$) time.
\item[Case 2.]
If $\OL{G}$ is unipolar, we find a maximum clique in $\OL{G}$. To do
this, we solve $k$ bipartite matching problems in $G$, and construct
a maximum independent set for the subgraph $G_{H \cup H_i}$ that has
the smallest maximum matching. The worst-case time for this is
O($n^{2.5}/\log n$).
\end{description}

Similarly, to find a maximum clique in a generalized split graph
$G$, we use a unipolar partition of $G$ or $\OL{G}$. If a unipolar
partition is not given as input, we use algorithm Unipolar\_Test to
obtain a unipolar partition of $\OL{G}$, if possible; otherwise, we
find a unipolar partition of $G$. In this case the time to find the
unipolar partition, which is O($nm + nm_F$) or O($n\OL{m} +
n\OL{m}_F$), can dominate the time to find a maximum clique.
\begin{description}
\item[Case 1.]
If $\OL{G}$ is unipolar, we find a maximum independent set in
$\OL{G}$. For each vertex $v$ of $H$, we can mark the cliques $H_i$
in which $v$ has a neighbor in $G$.  A scan of length no more than
the degree of $v$ plus 1 will determine if there is a clique $H_i$
in which $v$ has no neighbor in $G$.
Thus, this can be done in O($n+m$) time.
\item[Case 2.]
If $G$ is unipolar, we find a maximum clique in $G$. Here, we must
solve $k$ bipartite matching problems in $\OL{G}$, and construct a
maximum independent set for the subgraph $\OL{G}_{H \cup H_i}$ that
has the smallest maximum matching. The worst-case time for this is
O($n^{2.5}/\log n$).
\end{description}

\subsection{Minimum Clique Cover and Minimum Coloring}

Let $G=(V,E)$ be a unipolar graph that has a unipolar partition $H,
H_1, \ldots, H_k$. We first discuss finding a minimum coloring and
minimum clique cover of $G$. Then we present algorithms for
generalized split graphs.

Let $\alpha(G)$ be the size of a maximum independent set in $G$.
Finding a clique cover of $G$ of size $\alpha(G)$ is similar to
finding a maximum independent set in $G$. The cliques of the split
partition the vertices of $G$ into $k+1$ cliques. If there is a
vertex $x$ in $H$ that is non-adjacent in $G$ to at least one vertex
of each clique $H_i$, then $\alpha(G)=k+1$ and the unipolar
partition of $G$ is a minimum clique cover. Otherwise, $\alpha(G)=k$
and for every vertex $v$ of $H$ there is a clique $H_i$ such that
$\{v\} \cup H_i$ is a clique. A minimum clique cover of $G$ is
obtained by adding each vertex of $H$ to an appropriate peripheral
clique. These steps can be done in O($n+m$) time.

Let $\omega(G)$ be the size of a maximum clique in $G$. To produce a
proper coloring of $G$ using $\omega(G)$ colors we first properly
color each subgraph $G_{H \cup H_i}$ with $\omega(G_{H \cup H_i})$
colors. The maximum number of colors used in coloring any subgraph
is $\omega(G)$. So that the independent colorings of the subgraphs
agree on $H$ we first assign the vertices of $H$ the fixed distinct
colors {$1,2, \ldots, |H|$}. To color $G_{H \cup H_i}$ with
$\omega(G_{H \cup H_i})$ colors, we find a maximum matching $M$ in
$\OL{G}_{H \cup H_i}$, which is bipartite with bipartition $H, H_i$.
As discussed in Section~\ref{maxopt}, we have $|H \cup H_i| - |M| =
\omega(G_{H \cup H_i})$. Now we color both endpoints of each edge in
$M$ with the color preassigned to the endpoint in $H$. We have used
$|M|$ colors and $|H \cup H_i| - 2|M| = \omega(G_{H \cup H_i})-|M|$
vertices remain to be colored. Thus, each remaining vertex can
receive a distinct color; we use the preassigned colors on vertices
in $H$.  The running time for this algorithm is dominated by the
time to solve the $k$ bipartite matching problems in $\OL{G}$ and is
O($n^{2.5}/\log n$) (see Section~\ref{maxopt}).

To find a minimum clique cover in a generalized split graph $G$, we
use a unipolar partition of $G$ or $\OL{G}$. If a unipolar partition
is not given as input, we use algorithm Unipolar\_Test to obtain a
unipolar partition of $G$, if possible; otherwise, we find a
unipolar partition of $\OL{G}$. In this case the time to find the
unipolar partition, which is O($nm + nm_F$) or O($n\OL{m} +
n\OL{m}_F$), can dominate the time to find a minimum clique cover.
\begin{description}
\item[Case 1.]
If $G$ is unipolar, we find a minimum clique cover of $G$. This can
be done in O($n + m$) time.
\item[Case 2.]
If $\OL{G}$ is unipolar, we find a minimum coloring of $\OL{G}$. To
do this, we solve $k$ bipartite matching problems in $G$. The
worst-case time for this is O($n^{2.5}/\log n$).
\end{description}

To find a minimum coloring in a generalized split graph $G$, we use
a unipolar partition of $G$ or $\OL{G}$. If a unipolar partition is
not given as input, we use algorithm Unipolar\_Test to obtain a
unipolar partition of $\OL{G}$, if possible; otherwise, we find a
unipolar partition of $G$. In this case the time to find the
unipolar partition, which is O($nm + nm_F$) or O($n\OL{m} +
n\OL{m}_F$), can dominate the time to find a minimum coloring.
\begin{description}
\item[Case 1.]
If $\OL{G}$ is unipolar, we find a minimum clique cover of $\OL{G}$.
For each vertex $v$ of $H$, we can mark the cliques $H_i$ in which
$v$ has a neighbor in $G$.  A scan of length no more than the degree
of $v$ plus 1 will determine if there is a clique $H_i$ in which $v$
has no neighbor in $G$. Thus, this can be done in O($n + m$) time.
\item[Case 2.]
If $G$ is unipolar, we find a minimum coloring of $G$. Here, we must
solve $k$ bipartite matching problems in $\OL{G}$. The worst-case
time for this is O($n^{2.5}/\log n$).
\end{description}

\section{The Perfect Code Problem}

A {\it perfect code} in a finite, simple, undirected graph $G$ is an
independent subset $D \subseteq V$ such that each vertex $v \in V$
is either in $D$ or has exactly one neighbor in $D$. A perfect code
can be viewed as a partition of $V$ into the parts $\{N[v] \, | \, v
\in D\}$, where $N[v]$ is the closed neighborhood of $v$. Perfect
codes have been studied extensively (see \cite{haynes1}) and are
known by several other names: efficient dominating sets, $[0,
1]$-dominating sets, independent perfect dominating sets, and
$PDS_1$ sets.

Let Perfect Code denote the problem of
determining whether a given graph contains a perfect code.
Perfect Code is easily solved for split graphs.
Let $G=(K \cup I,E)$ be a split graph whose vertex set is partitioned into clique $K$ and
independent set $I$.  A perfect code contains a vertex $x$ of $K$ only if each vertex
of $I$ is either adjacent to $x$ or an isolated vertex.  If a perfect code does not contain
a vertex of $K$, then the neighborhoods of the vertices of $I$ must partition $K$.
See \cite{mchang} for algorithms for the weighted version of the problem
on split and chordal graphs.
There are also efficient algorithms for
numerous other graph classes: series-parallel graphs \cite{amin2},
interval and circular-arc graphs \cite{wfk-ee}, and
others (see, e.g., \cite{chang, stout1}).

Perfect Code is NP-Complete for general graphs (and remains
NP-Complete for bipartite graphs, chordal graphs, and planar graphs
of maximum degree 3); several such proofs have appeared in the literature
(see, e.g., \cite{BBS, fellows, krat1, krat2, smart, yen}).
This implies that Perfect Code is NP-Complete for co-unipolar and
generalized split graphs, since every bipartite graph is
co-unipolar.

Smart and Slater \cite{haynes1, smart} gave simple reductions from
the NP-Complete problem One-in-Three 3SAT (with no negated
variables) (see \cite{GJ}) to prove that Perfect Code is NP-Complete
for bipartite and chordal graphs.  The reduction for the chordal
case produces a graph that is also unipolar, thus establishing that
Perfect Code is NP-Complete for chordal unipolar graphs.  To the
best of our knowledge, this is the only problem shown to be
NP-complete on unipolar graphs.

\noindent {\bf Acknowledgement} The authors thank John Goldwasser
for his suggestion on the analysis of the maximum clique algorithm
for unipolar graphs.


\end{document}